\documentclass[amsfonts, amssymb, amsmath, prfluids, showkeys, nofootinbib, superscriptaddress,twoside]{revtex4-2}
\usepackage[english]{babel}
\usepackage[utf8]{inputenc}
\usepackage{amsthm}
\usepackage{mathtools}
\usepackage{physics}
\usepackage{xcolor}
\usepackage{graphicx}
\usepackage[normalem]{ulem}
\usepackage[left=23mm,right=13mm,top=35mm,columnsep=15pt]{geometry} 
\graphicspath{ {./figures/} }
\usepackage{adjustbox}
\usepackage{placeins}
\usepackage[T1]{fontenc}
\definecolor{colortodo}{RGB}{0,250,0} 

\usepackage[colorinlistoftodos, color=green!40,prependcaption]{todonotes}
\usepackage[pdftex, pdftitle={Article}, pdfauthor={Author}]{hyperref} 
\begin{document}

\title{Alcove formation in dissolving cliffs driven by density inversion instability}
\author{Ram Sudhir Sharma}
\affiliation{Department of Physics, Clark University, Worcester, MA 01610, USA}
\author{Michael Berhanu}
\affiliation{MSC, Universit\'e de Paris, CNRS (UMR 7057), 75013 Paris, France}
\author{Arshad Kudrolli}
\email{akudrolli@clarku.edu}
\affiliation{Department of Physics, Clark University, Worcester, MA 01610, USA}

\date{\today} 

\begin{abstract}
We demonstrate conditions that give rise to cave-like features commonly found in dissolving cliffsides with a minimal two-phase physical model.  Alcoves that are wider at the top and tapered at the bottom, with sharp-edged ceilings and sloping floors, are shown to develop on vertical solid surfaces dissolving in aqueous solutions. As evident from descending plumes, sufficiently large indentations evolve into alcoves as a result of the faster dissolution of the ceiling due to a solutal Rayleigh-B\'enard density inversion instability. By contrast, defects of size below the boundary layer thickness set by the  critical Rayleigh number smooth out, leading to stable planar interfaces. The ceiling recession rate and the alcove opening area evolution are shown to be given to first order by the critical Rayleigh number.  By tracking passive tracers in the fluid phase, we show that the alcoves are shaped by the detachment of the boundary layer flow and the appearance of a pinned vortex at the leading edge of the indentations. The attached boundary layer past the developing alcove is then found {\color{black} to} lead to rounding of the other sides and the gradual sloping of the floor. 
\end{abstract}


\maketitle

\section{Introduction} \label{sec:introduction}
The coupled evolution of solid-fluid interfaces due to dissolution, melting, and erosion is important in shaping geophysical features including mountains, icebergs, caverns, aquifers, and petroleum reservoirs~\cite{Fredd1998,Ortoleva1994,Malin2000,Meakin2010}. While typically slow, such evolution can lead to dramatic appearance of sinkholes, avalanches, and other modes of rapid failure. There is considerable need to understand the evolution of such features from a mechanistic perspective  
to complement field observations towards developing quantitative models because of their complexity and ubiquity both on Earth and other celestial bodies~\cite{Malin2000,Kudrolli2016,Jerolmack2019}.

Irregular surface patterns in salt crystals dissolving from below in aqueous solutions have been observed and analyzed with turbulent boundary layer models~\cite{Thomas1968,Sullivan96}.   Differential growth resulting in up-facing conical cavities created within salt deposits shaped by dissolution driven internal flows have been investigated in the context of land subsidence and structure collapse~\cite{Gechter2008}. Opacity of the medium only allowed intermittent and indirect inferences to be drawn in these studies.
Recent studies of two-phase model systems with rapidly dissolving non-crystalline hard candy have illuminated the formation of pinnacles in karsts, furrows in sandstone, and ice scallops~\cite{Huang2015,Nakouzi2015,Cohen2016,Huang2020,Wykes2018,Pegler2020,Cohen2020,bushuk_holland_stanton_stern_gray_2019}. Significant progress has been made in describing the observed outer envelops of the structures with counter-intuitive growth of singularities versus blunting of sharp tips depending on the shapes of the initial surfaces and flow conditions~\cite{Pegler2020}. While occurrence of surface cavities were noted as an undesirable outcome in their studies, they were not explored.  Moreover, cavity growth is of interest in an even broader class of systems, as in fluid-mediated pitting corrosion which leads to deep isolated holes and failure in metal structures. 

\begin{figure}
\begin{center}
\includegraphics[width=.9\linewidth]{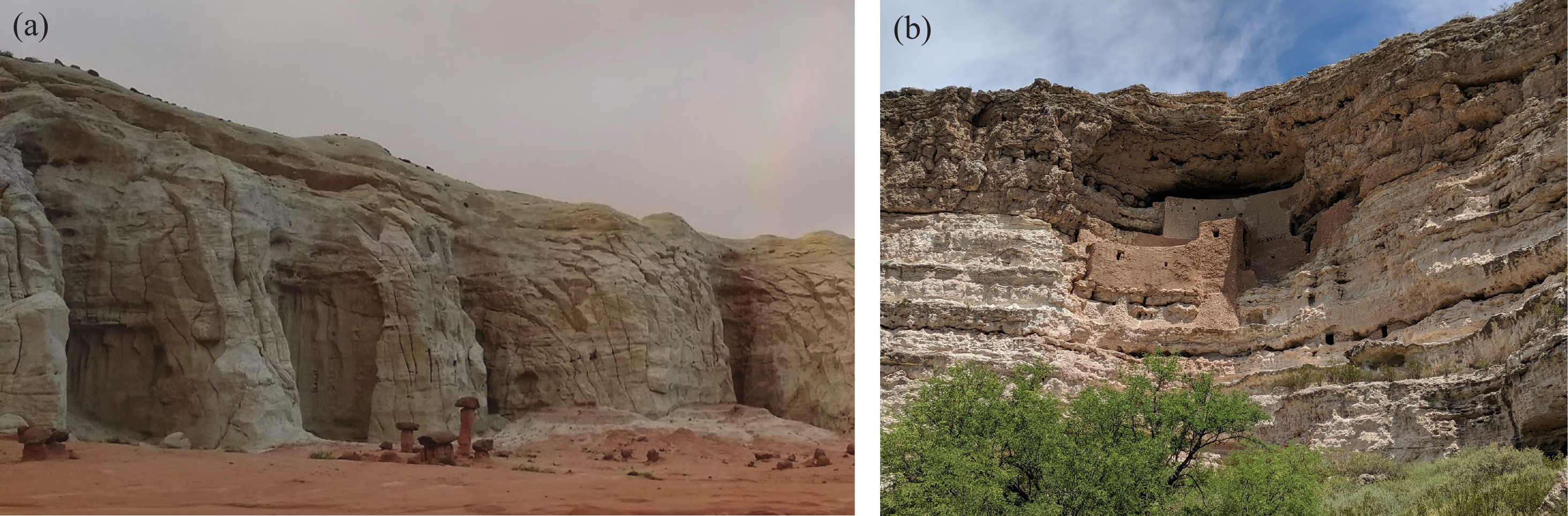}
\caption{Examples of alcoves observed in 
(a) Grand Staircase-Escalante National Monument, Utah, and (b) Montezuma Castle National Monument, Arizona which has pre-Columbian dwellings built within. The cliffs are approximately 20 and 30 meters high for scale, respectively.  Photos taken on June 21, 2019 and June 24, 2019, respectively. 
}
\label{fig:alcove}
\end{center}
\end{figure}

Here, we focus on the evolution of vertical solid-liquid interfaces as a result of coupled fluid flow and dissolution of a solid. Figure~\ref{fig:alcove} shows motivational examples of alcoves - cave-like features that occur above ground on cliffs - found in the American Southwest and thought to result from fluid flow and weathering.  Dissolution of calcium carbonate, and the repeated freezing and expansion cycles which weaken and loosen pieces are known to lead to alcoves within the cliffs as they erode. However, the minimal conditions under which they can occur and the detailed mechanism by which they evolve beyond these general descriptions remain unclear. 

Therefore, we investigate if alcoves can be produced by using a highly simplified two-phase model composed of a homogeneous noncrystalline solid and a dissolving fluid phase in a gravitational field. The relatively high transparency and dissolution speed of the medium enables us to obtain detailed evolution information on the of shapes with complementary optical and x-ray measurements, and the flow fields in a matter of hours if not a day. Specifically, we show perturbations in a vertical dissolving interface develop into alcoves shaped by the interactions with the surrounding fluid as a result of a density inversion instability at the underside of surface perturbations. In contrast, up-facing horizontal surfaces are observed to dissolve more uniformly and much more slowly, even when surface indentations and material defects are present, under otherwise similar conditions.

\section{Materials and Methods}
The solids used in our study are prepared with 5 part sucrose, 3 part light corn syrup, and 2 part distilled water 
by volume, similar to recipes in other studies~\cite{Huang2015,Huang2020}. These ingredients are stirred and heated together, while keeping the temperature at approximately $160^\circ$C, {\color{black} until} 7/8th of the original volume remains after evaporation corresponding to a density $1.3 \pm 0.1$ g/cm$^3$. The {\color{black} hot liquid} is poured into a 3D printed rectangular transparent mold with internal width $W$, length $L$, and height $H$. Typically, we use $W =  4$\,cm, length $L=6$\,cm, and height $H=0.3$\,cm to standardize our measurements, and larger molds to check any mold dimension dependence.  The {\color{black} hot liquid} solidifies with a flat surface in gravity, with prescribed surface indentations imposed  as needed during the final stages of cooling as the glass transition is approached.  Air bubbles from submicron to a few millimeters in size generated during the mixing and heating process can be trapped in, and on, the surface of the {\color{black} hot liquid} at it cools. The larger visible bubbles were eliminated by trial and error, and degassing as needed. Surface indentations are also added if desired as the liquid hardens using solid impalers with prescribed shapes. Using this preparation method, we obtain a homogeneous noncrystalline solid with a uniform shade of color, and density $\rho_s = 1.41 \pm 0.06$\,g/cm$^3$. 

The mold with solid within is then placed in a transparent acrylic rectangular container filled with distilled water and oriented as desired with one surface exposed.  A surfactant is added to reduce air bubbles in the fluid. 
A sufficiently large bath is used so that density of the bath solution even after all the solid dissolves varied {\color{black} by} less than 0.5\%. The experiments were performed in a laboratory with temperature $T$ in the range $21^\circ$C to $25.1^\circ$C as noted. The solid is imaged with a megapixel Pixelink color camera through the sidewalls of the container by back-lighting with a uniform LED panel. The image intensity is mapped to the thickness of the solid enabling us to obtain a map of the surface dynamically (see Appendix~\ref{sec:img}). 
The optical measurements are also complemented with snapshots of the solid with a Varian Medical Systems micro x-ray CT instrument. This requires us to pull the solid and the mold out of the bath, and was mostly conducted to illustrate and check the overall shape of partially dissolved solids.

\section{Observation of Alcove Growth} 

\begin{figure}
\begin{center}
\includegraphics[width=17cm]{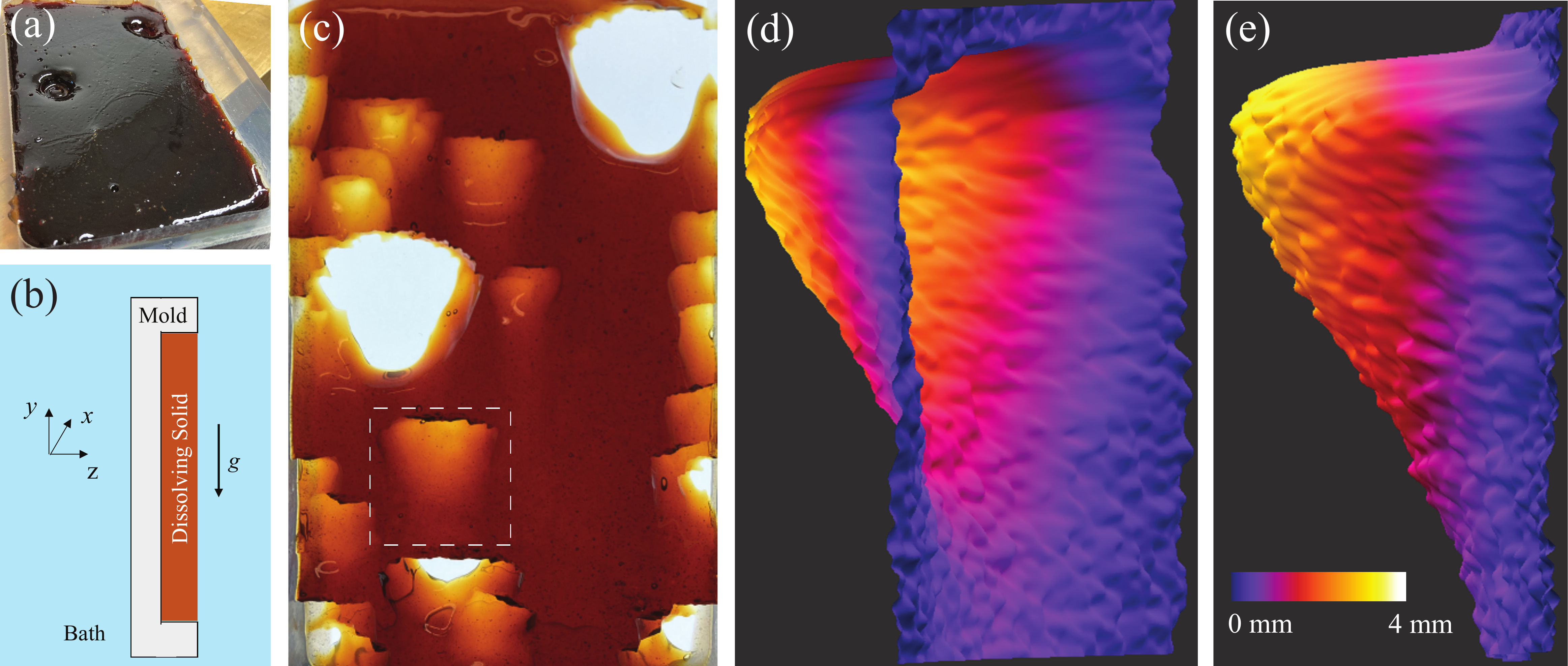}
\caption{(a) Image of solid sugar block with two circular indentations contained within a transparent mold. ($L=14$\,cm; $W=9$\,cm; $H=1$\,cm). Several other small and large defects are also made visible by reflected light. (b) Schematic crosssection of the sugar block contained within the mold placed vertically in an aqueous bath and Cartesian coordinate system. (c) The surface shows development of a number of alcoves after being immersed for an hour. 
(d) A surface rendering obtained with x-ray scanning displayed at a $60^\circ$. The alcove shown corresponds to the one marked with dashed box corresponding to a 3\,cm by 3\,cm area in (c). (e) Same surface plot shown from side.  
}
\label{fig:intro}
\end{center}
\end{figure}

Figure~\ref{fig:intro}(a) shows an example of a surface prepared with two large circular indentations besides a number of smaller surface imperfections formed during the molding process.  The solid and the mold are immersed while oriented vertically in the center of a  bath container for 60\,minutes as shown schematically in Fig.~\ref{fig:intro}(b). An image of the partially dissolved solid where a number of pits or alcove-like features have appeared is shown in Fig.~\ref{fig:intro}(c). Comparing Fig.~\ref{fig:intro}(a) and Fig.~\ref{fig:intro}(c), the largest alcoves correspond to the two initial circular indentations, but many other smaller ones are also observed. A surface rendering of a typical alcove obtained with x-ray scanning - corresponding to the region indicated by Fig.~\ref{fig:intro}(c) - is shown in Fig.~\ref{fig:intro}(d) and (e) as viewed at a 60$^\circ$ angle and from the side, respectively.  A sharp transition from the vertical interface to the ceiling can be observed, whereas the floor is very sloped and meets the vertical interface with a further gradual change of slope. The alcoves are wider near the top and appear conical below, giving rise to an overall inverted triangle appearance with flat ceilings, reminiscent of the alcoves seen in Fig.~\ref{fig:alcove}.

\begin{figure}
\begin{center}
\includegraphics[width=18cm]{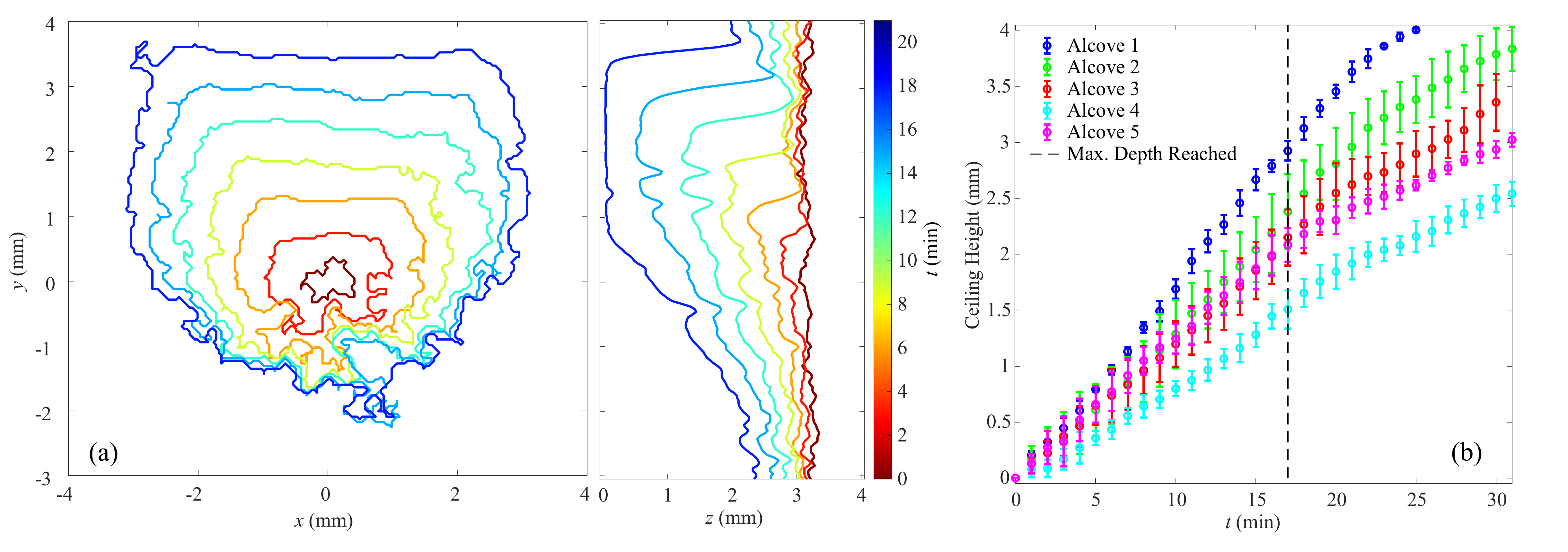}
\caption{(a) The alcove outline and vertical mid-profile shows the evolution of the ceiling and floor. {\color{black} The outline correspond to a contour $h = 0.1H$ from the vertical surface, and its roughness is predominantly due to imaging artifacts. } (b) The average change in ceiling height as a function time based on 5 spontaneously formed alcoves under similar conditions. The ceiling is measured to recede up at a constant rate range between $\dot{\eta}$ = 6.0 mm/hr and $\dot{\eta}$ = 10.7 mm/hr till the back wall is reached (vertical dashed line) before slowing down.  The measured rate averaged over the 5 examples $\dot{\eta}$ = 8.4 mm/hr is consistent with the estimated rate 9.4 mm/hr from Eq.~(\ref{eq:rrates}) corresponding to a solutal Rayleigh-B\'enard instability.}
\label{fig:time}
\end{center}
\end{figure}

From a snapshot alone it is difficult to match each of the alcoves with a defect, but it is evident that defects play an important part in their formation. Therefore, we performed complementary experiments with a 6 cm by 4 cm by 0.3 cm solid with an initial 0.1\,cm fluid filled gap between the dissolving interface and container boundary to enable clear visualization. Figure~\ref{fig:time}(a) shows the evolution of the outline and vertical transect of an alcove observed using optical imaging at various stages of development. {\color{black} The calibration used to map the image intensity to the local surface height can be found in Appendix~\ref{app:smooth}.} The corresponding movie can be found in the supplementary documentation~\cite{sup-doc}. Here, an essentially flat surface is observed to develop an alcove which grows longer, wider, and deeper until it encounters the back wall of the mold. Even beyond that point, the alcove is observed to continue to grow longer and wider. Thus, the alcove is observed to evolve {\color{black} to have an inverted triangle appearance} with a sharp transition to the horizontal roof from the vertical face, and a smooth sloping floor which gradually transitions to the vertical surface similar to that shown in Fig.~\ref{fig:intro}(d,e).

This evolution indicates that all the alcove-like features observed in Fig.~\ref{fig:intro}(c) are similar in nature, and that the smaller ones start to form later as the dissolving interface reaches small air bubbles trapped inside the dissolving solid. While some surface indentations develop into alcoves, adjacent vertical regions become optically smooth compared to even the initial surface which may have small pits and grooves on the solid surface formed during the cooling process. Thus, it appears that not all surface perturbations become unstable, and the overall appearance of the isolated alcoves is different than the irregular scallops which develop over a surface melting or dissolving from below~\cite{Sullivan96,Meakin2010}. 

Further comparing the crosssections at different time intervals, we observe that the ceiling recedes the fastest compared to all the other surfaces. We measure the change in the ceiling height as a function of time $t$ for 5 different examples of spontaneously formed alcoves in our experiments (see Fig.~\ref{fig:time}(b)). While there is some variation, the time dependence is more or less similar with essentially linear growth before a slowing down once the alcove reaches the back-wall of the mold {\color{black} at $16.3\,{\rm min} \pm 0.3$\,min in the various cases}.  We thus postulate that alcoves start to develop at inverted regions corresponding to indentations, or bubble defects with a critical size, when the ceilings recede rapidly and spread horizontally as the solid dissolves.

\section{Boundary Layer Analysis} 

\begin{figure}
\centering
\includegraphics[width=0.95\textwidth]{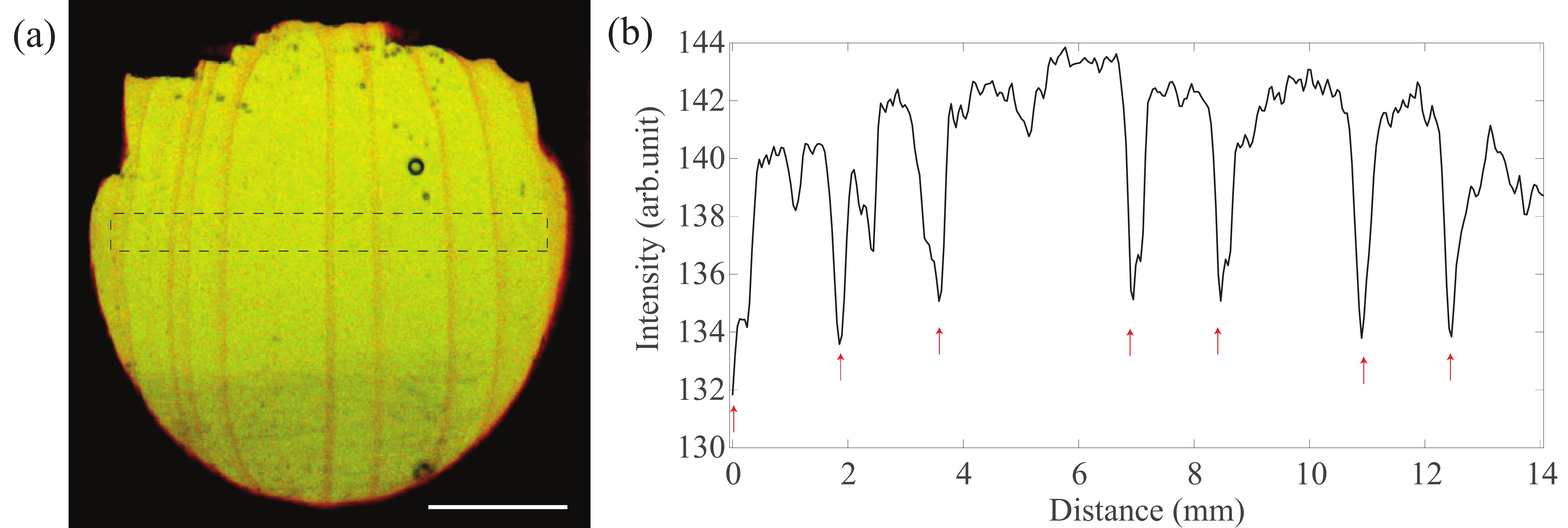}
\caption{(a) Image of plumes descending from the ceiling of an alcove ($t= 21$\,min). {\color{black} A few stray air bubbles which appear as small circles are also visible. The scale bar corresponds to 5\,mm. } 
(b) Intensity profiles as a function of horizontal distance corresponding to dashed box shown in (a) with locations of plumes indicated by arrows. Average plume spacing is observed to be approximately 2\,mm consistent with critical wavelength $\lambda_c$ estimate corresponding to solutal Rayleigh-B\'enard instability. 
}
\label{fig:plume}
\end{figure}

When a solid is immersed in a solvent bath, a concentration gradient of the solute develops at the interface, forming a boundary layer, which grows by diffusion~\cite{Garner1961}. This boundary layer is stable in gravity when the dissolving interface is below the solution and the solution density increases with solute concentration. If the dissolving surface is tilted or of finite size, boundary layer flows can develop along the slope under the action of the gravity, but remain attached to the solid interface. Such conditions shape the dissolving body at large scales~\cite{Wykes2018,Pegler2020,Pegler2021}. In contrast, if the solid is above the solvent bath, a density inversion occurs rapidly as the boundary layer becomes more dense than the solvent below. 
Thus, gravity can destabilize the boundary layer generating descending plumes similar to the miscible Rayleigh-Taylor instability~\cite{Chandrasekhar}. {\color{black} Because diffusion and the fluid viscosity can oppose the destabilizing effect of gravity, the concentrated boundary layer becomes unstable only when its thickness exceeds a critical value.} Then, the instability is more accurately identified with the solutal Rayleigh-Bénard instability~\cite{Sullivan96,Philippi2019}. The boundary layer is continuously regenerated by dissolution between the emission of successive plumes, so that its thickness remains on average close to the critical value, as shown experimentally~\cite{Sullivan96,Wykes2018,Cohen2020} and numerically~\cite{Philippi2019}. As we argue and demonstrate in the following, this analysis developed on the dissolution of inverted surfaces can be applied here to the ceiling of the alcoves.

While it is not possible for us to directly show the boundary layer, we can visualize the descending plumes when they can be contrasted with a bright background. This is possible by using an alcove with a clear back-wall corresponding to that of the mold. As shown in Fig.~\ref{fig:plume}(a), faint plumes descending from the ceiling can be directly observed in the closeup image and in the associated movie in the supplementary documentation~\cite{sup-doc}. Plotting the average intensity across the region marked by the dashed box in Fig.~\ref{fig:plume}(a), we observe 7-8 plumes clearly visible corresponding to an approximate spacing of 2 \,mm. Similar plumes have been observed descending from spherical candy blocks as well giving rise to nonuniform dissolution of the solid with a smooth top-half surface, and a flatter bottom with sharp edges due to flow separation~\cite{Wykes2018}.   These plumes rapidly take away the dense solution from the boundary allowing fresh fluid to enter the region, and allowing further dissolution to occur. Thus, this instability allows a relatively rapid recession of inverted dissolving boundaries to occur relative to other orientations as noted in Fig.~\ref{fig:time}(a). 

The critical thickness of the concentration boundary layer $\delta_c$ that can cause the solutal Rayleigh-Bénard instability between a rigid boundary and a stress-free boundary is given by~\cite{Sullivan96,Wykes2018}:
\begin{equation}
\delta_c = \left(\frac{{\rm Ra}_c \, \mu \, \rho_f \, D}{g \Delta \,\rho}\right)^{1/3}, 
\label{Eq1}
\end{equation}
where Ra$_c = 1101$ is the critical Rayleigh number corresponding to rigid-free mixed boundary conditions~\cite{Chandrasekhar}, $g = 9.8$\,m/s$^2$ is the gravitational acceleration, $\Delta \rho$ the density difference across the layer, $\mu$ the kinematic viscosity, $\rho_f$ the density of the far-field fluid, and $D$ the molecular diffusion constant of the dissolved species.  During the entire dissolution process, the thickness of the concentration boundary layer remains close to its critical value $\delta_c$ due to the emission of the plumes.
Because the solid-fluid material system is similar to those used in Ref.~\cite{Wykes2018}, we follow estimates there that $\Delta \rho = 300$ kg/m$^3$, and $D = 4.3 \times 10^{-10}$m$^2$/s. Then, assuming a viscosity of the sugar solution averaged between the saturation value ($\mu = 7.7 \times 10^{-4}$\,m$^2$/s) next to the solid, and water ($\mu = 1.0 \times 10^{-6}$\,m$^2$/s) outside the boundary layer, we find $\delta_c \approx 0.40$\,mm. The Schmidt number is used to estimate the relative magnitude of momentum diffusivity to mass diffusivity, and is given by ${\rm Sc} = \mu/D$. 
{\color{black} Substituting in the viscosity corresponding to water to obtain a lower bound,} we have Sc $\approx 10^3$. Hence, viscosity dominates diffusion and the solute can be expected to be confined within the boundary layer. Under these conditions, the critical wavelength of the instability $\lambda_c \approx 5\, \delta_c = 2$\,mm. This estimate is remarkably consistent with the mean plume spacing seen in the example shown in Fig.~\ref{fig:plume}. Thus, $\delta_c$ can provide a critical length scale for an inverted region needed to trigger a flow instability which can lead to enhanced dissolution. Conversely, smaller perturbations may not be sufficient for the instability to develop and would thus smooth out over time as the surface dissolves stably.   

The recession rate in terms of the location of the interface $\eta$ relative to the initial surface is obtained by writing two conservation laws at the interface. The conservation of the mass:
$$-\rho_s \frac{d\eta}{dt}=\rho_l\,\left(\mathbf{u_i}\cdot \mathbf{n} -  \frac{d\eta}{dt}\right)\, ,$$ with $\mathbf{u_i}$ the fluid velocity at the interface, $ \mathbf{n}$ a unitary vector normal to the interface and $\rho_l$ the density of the liquid at the interface. Then, the conservation of the solute gives:
$$\rho_s \frac{d\eta}{dt}=\left(\frac{d\eta}{dt}-\mathbf{u_i}\cdot \mathbf{n} \right)\, c_i + D\,\mathbf{\nabla} c \cdot \mathbf{n}\, , $$ where $c$ is the mass concentration of the solute in the solution, and $c_i$ is the concentration of the solute at the interface.
The last term corresponds to the diffusive flux at the interface according to the Fick's law. For fast dissolving species, the interface concentration $c_i$ is, to a good approximation, very close to the saturation concentration $c_{sat}$~\cite{Philippi2019}, and thus the liquid density $\rho_l$ is also close to the saturation density $\rho_{sat}$ . The diffusive flux $D\,\mathbf{\nabla} c \cdot \mathbf{n}$ can be then approximated by $D\,c_{sat}/\delta$ if the bath can be approximated as fresh water. By combining these last two equations, we obtain:
\begin{equation}
\frac{d\eta}{dt} = \frac{D \, c_{sat}}{\rho_s \, \delta \, (1-c_{sat}/\rho_{sat})}\,,
    \label{eq:rrates}
\end{equation}
where 
the saturation concentration of sucrose $c_ {sat} =0.67$, $\rho_s=940$\,kg/m$^3$, and the liquid density at saturation $\rho_{sat} = 1300$ kg/m$^3$~\cite{Wykes2018}, while starting with distilled water as the solvent. 

Plugging in the diffusion and concentrations corresponding to our experiments,  we estimate the solutal Rayleigh-Bénard instability implied recession rate  $\dot{\eta} = {d\eta}/{dt} = 9.4$\,mm/hr by then using $\delta = \delta_c$. This {\color{black} estimated value of $\dot{\eta}$} is remarkably consistent with the measured rate $\approx 8.4$\,mm/hr at which the ceiling recedes up upward in the examples shown in Fig.~\ref{fig:time}(b). The agreement observed with the plumes and the recession rate thus clearly identifies the nature of the instability leading to alcoves as being consistent with solutal Rayleigh B\'enard instability in our experiments.

\section{Alcove Evolution}
\begin{figure}
\centering
\includegraphics[width=0.95\textwidth]{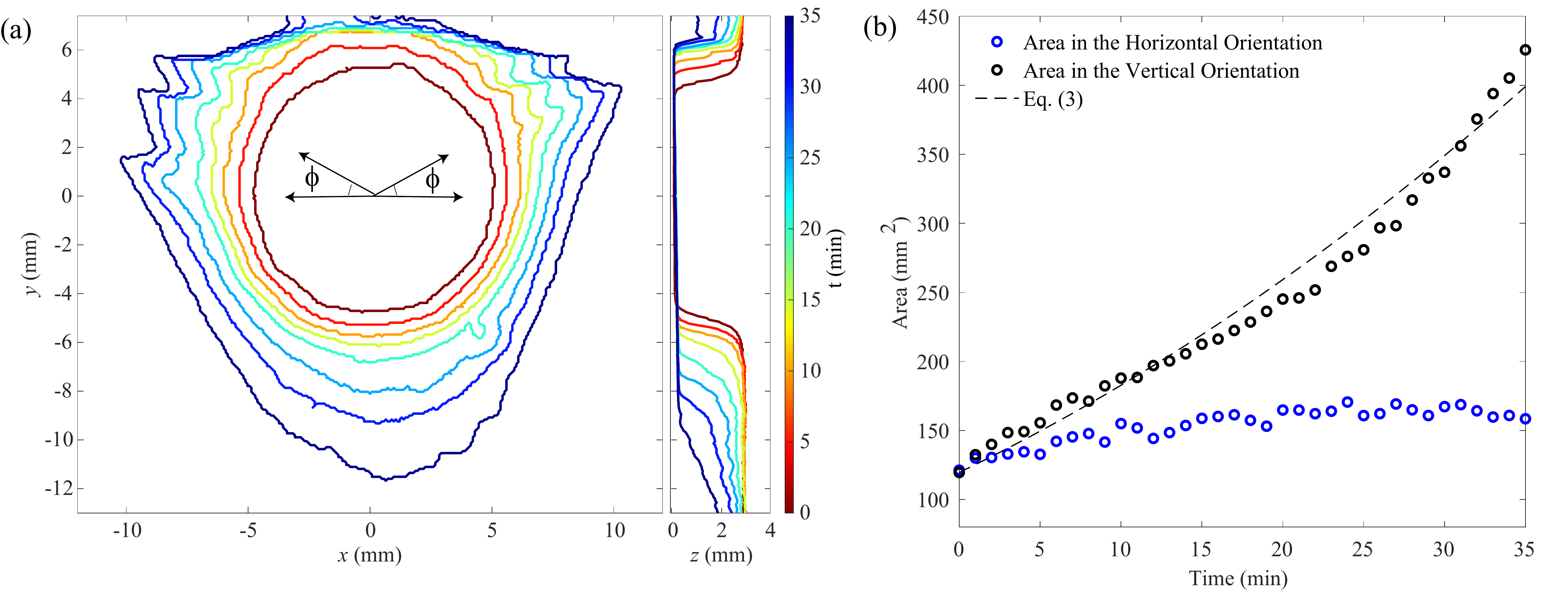}
\caption{(a) Contour corresponding to $h = 0.1H$ to show the progression of the alcove opening shape and vertical crosssection. {\color{black} The angle $\phi = - 20 \pm 5^0$ and $20 \pm 5^0$ at which triangular corners develop are also shown.} (b) The area contained with the contour for the case shown vertically, and when a similar initial interface is placed horizontally (see Appendix~\ref{sec:hori}). Dissolution progresses much more rapidly when the interface is vertical. {\color{black} A fit to Eq.~(\ref{eq:area}) with $K =  1.647$, 95\% confidence bounds (1.609, 1.685), and $R_o = 6.17$\,mm is observed to capture the overall increase in alcove opening area in the vertical case  surprising well considering that a circular opening is assumed}.   
}
\label{fig:vertical}
\end{figure}

{\color{black} Having established predictions for} the conditions under which the alcoves initiate, we next examine the evolution of the alcove opening shape. We simplify and standardize the surface perturbation that lead to alcoves by starting with a cylindrical cavity of radius $R_o = 5$\,mm in the center of $L = 60$\,mm by $W = 40$\,mm by $H = 3$\,mm mold, while minimizing the bubble defects in the solid. We place the vertical interface a distance $d_{gap} = 1$\,mm parallel to a sidewall of the bath container to simplify imaging, and distilled water is then added to start the dissolution. As long as fluid is allowed to pass across and $d_{gap}$ is greater than the boundary layer flow, the relative location of the dissolving interface in relation to a sidewall is found to have no particular effect on the phenomena of interest. 

Figure~\ref{fig:vertical}(a) shows the contours in the $x-y$ plane corresponding to a depth of $0.1H$ from the vertical surface in 240\,s time intervals to quantify the development of the alcove opening. We observe that the ceiling of the cavity flattens over time, while the bottom tapers and becomes elongated as it grows to resemble the ones seen in Fig.~\ref{fig:intro}(c) and Fig.~\ref{fig:time}(a).  Measured from the initial center ($x=y=0$\,cm), {\color{black} the alcove opening develops peaks or triangular corners at approximately $\phi = -20^\circ$ and $\phi = 20^\circ$ relative to the $x$-axis}. Thus, the shape grows rather symmetrically about the flow direction considering the imperfections in the solids.  Some imperfections in the shape of the initial circular shape and a few air bubbles can be noted in the images as dissolution progresses. But, these imperfections appear to not affect the evolution of the shape downstream, and the overall development of the alcove. 

\begin{figure}
\begin{center}
\includegraphics[width=7cm]{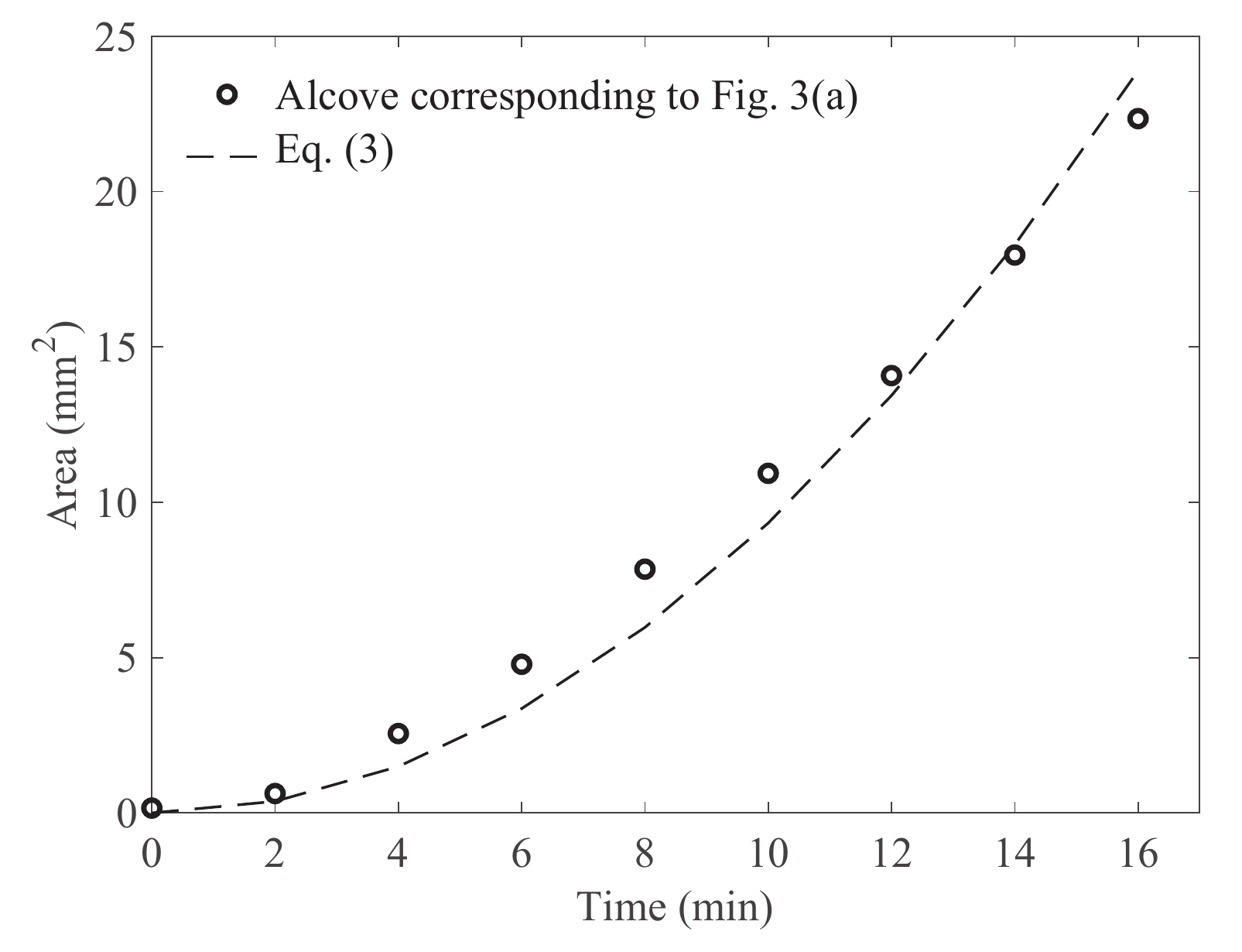}
\caption{{\color{black} The alcove opening area corresponding to the example shown in Fig.~\ref{fig:time}(a). Eq.~(\ref{eq:area}) with $K = 1.104$, 95\% confidence bounds $(1.073, 1.137)$, and $R_o = 0$\,mm is observed to describe the area evolution. } } 
\label{fig:area2}
\end{center}
\end{figure}

Fig.~\ref{fig:vertical}(b) shows the area of the alcove opening $A(t)$ enclosed by the contours corresponding to a small depth $h = 0.1H$ from the vertical surface. We observe that $A(t)$ grows increasingly rapidly with time. Further comparing the opening area of the horizontal cavity as a function of time in Fig.~\ref{fig:vertical}(b), we observe that it does not grow significantly over the same time interval (see Appendix~\ref{sec:hori}). While not circular, the alcove grows in all directions. 
{\color{black}The recession rate $\dot{\eta}$ given by Eq.~(\ref{eq:rrates}) is only strictly valid at the inverted alcove ceiling. Nonetheless, we assume that $\dot{\eta}$ provides a first order estimate of the magnitude of the gravity driven dissolution rate all along the alcove contour (see Fig.~\ref{fig:time}(a)). Then, the opening area of the alcove  $A(t)$ as a function of time $t$ can be estimated as},  
\begin{equation}
    A(t) = \pi (R_o + K \dot{\eta}\, t)^2, 
    \label{eq:area}
\end{equation}
where $R_o$ is the radius of the indentation at time $t=0$, and $K$ is a geometric fitting parameter of O(1). This measured area of the alcove enclosed by the contours is plotted along with the fits to Eq.~(\ref{eq:area}) in Fig.~\ref{fig:vertical}(b). 
{\color{black} The fit 
appears reasonable with $K \approx 1.65$, and $R_o = 6.17$\,mm implying that the average contour expansion speed is essentially constant. (A somewhat larger $R_o$ fit-value is used compared to initially prepared 5\,mm circular radius because the surface starts to dissolve as soon as the mold is placed in the water bath before the imaging starts.) Averaging further over four examples of alcove growth, we find $K$ in the range 1.44 and 2.77. We also compare Eq.~(\ref{eq:area}) with area evolution corresponding to the spontaneously formed alcove shown in Fig.~\ref{fig:time}(a). As can be seen from Fig.~\ref{fig:area2}, we observe good agreement with $K \approx 1.1$ and $R_o = 0$\,cm, since the initial size of a spontaneously formed alcove is negligible. Therefore, $\dot{\eta}$ given by Eq.~(\ref{eq:rrates}) gives a surprisingly good estimate of the overall alcove opening formed under different conditions, even though the hypothesis of uniform expansion does not apply to the contours at all times. 
A systematic overestimate may be expected because the model assumes that the alcove remains cylindrical, whereas the floors are sloped significantly giving rise to a larger area while considering the contour given by $0.1H$.} One observes from the central crosssectional view shown in Fig.~\ref{fig:vertical}(a) that the transition upstream from the vertical interface above the alcove to the ceiling remains sharp even as it retreats upwards. Whereas, the floor begins to slope and the edge round out, leading to a gradual transition downstream over time.

The evolution of the floor slope may be viewed as being similar to those reported in the development of pinnacles~\cite{Huang2020,Pegler2020} as when an attached flow shapes dissolution over large scales. There it was found that the pinnacle tip {\color{black} descends at a rate proportional to the curvature of the tip to the 1/4 power.} 
{\color{black} However, the overall geometries are different because of the  convex tip shape in the case of the pinnacles, and the  concave shape at the base of the alcove. A more recent theoretical study in two-dimensions~\cite{Pegler2021}  addresses the shape evolution of a soluble body with an attached boundary layer flow. The predicted shape evolution is qualitatively similar to the floor change plotted in Fig.~\ref{fig:vertical}(a) in the $y-z$ plane. Nonetheless, this model does not capture the triangle shape observed experimentally in the $x-y$ plane. } Assuming that the thickness of the solid block is the appropriate length scale in our system, we note a faster retreat of the floor downwards, as the solid thickness decreases (see Fig.~\ref{fig:vertical}(a)). In contrast, the retreat of the floor downwards slows down and even stops over time in the example shown in Fig.~\ref{fig:time}(a), as the alcove depth increases starting from a small perturbation.

\section{Gravity current and boundary separation}
\label{sec:flow}

\begin{figure}
\centering
\includegraphics[width=0.95\textwidth]{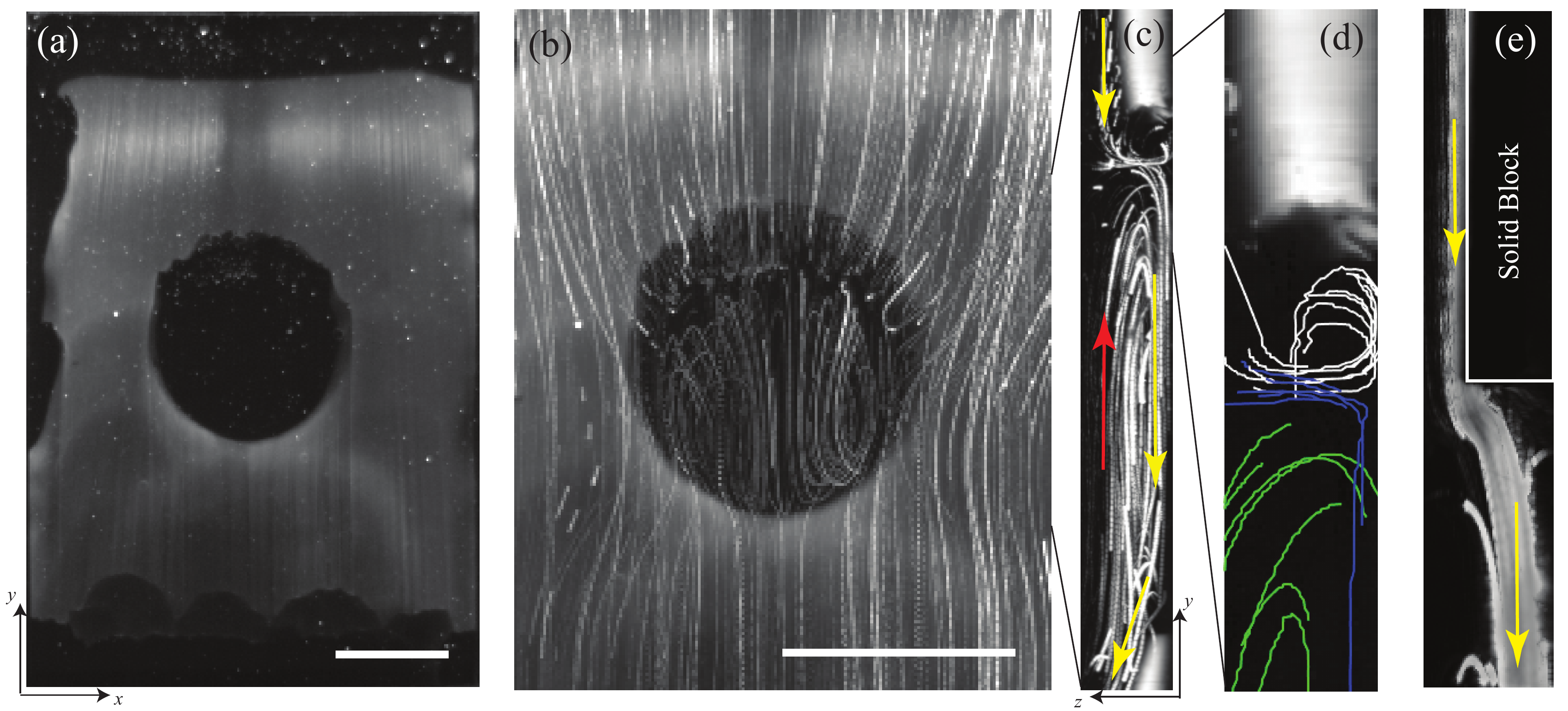}
\caption{(a) An image of the partially dissolved surface contained within a 6\,cm by 4\,cm vertically oriented mold at time $t = 20$\,min with fluorescent tracers added to visualize flow. (b) A zoomed in view of the fluid flow illustrated by using tracks of the tracers over a $\Delta t = 2$\,s time period.  A gravity driven downward current is observed along the dissolving surface. Circulation can be also observed inside the alcove.   (c) A central alcove crosssection showing the circulating currents within the alcove. The flow is observed to detach from the edge where the vertical dissolving surface meets the alcove ceiling. (d) A magnified view of the region containing the counterclockwise rotating vortex observed near the ceiling, the flow reattachment, and the clockwise rotating vortex further downstream with tracer tracks denoted with white, blue, and green colors, respectively. (e) A $\Delta t = 2$\,s long exposure image processed with maximum intensity and background subtraction {\color{black} reveals} the dominant flow path for the gravity current. The white bar in (a,b) corresponds to 1\,cm, and the width of the sugar block in (c-d) corresponds to 3\,mm for scale.}
\label{fig:flow}
\end{figure}

We investigate the impact of the fluid flow on the shaping of the dissolving interface by adding micro-sphere tracers to the solution using a laser sheet  with wavelength 532\,nm  to uniformly illuminate the area of interest.  The latex spheres with diameter $d_{ms} = 15\,\mu$m and density $\rho_{ms} = 1.1$\,g/cm$^3$  fluoresce and a bandpass filter is placed in front of the camera to block directly reflected light from entering the camera. This enables us to capture an image of the tracers with higher degree of contrast compared to the background. Sucrose was added to the initial bath to match that density while doing these measurements to reduce any settling due to density differences between the tracers and the fluid.

Figure~\ref{fig:flow}(a) shows an image from the front while de-focusing the laser sheet and illuminating the entire $d_{gap}=1$\,mm fluid gap between the dissolving interface and the front boundary from the side.  Figure~\ref{fig:flow}(b) then shows streaks made by the tracers obtained by adding successive images over 2\,s long using the maximum intensity function in ImageJ. The flow can be observed to be more or less uniformly downward outside the alcove but also converges weakly and then diverges as it passes the alcove. Further, flows in and out of the projected plane are also clearly visible inside the alcove. 
Thus, it can be noted that while the flow is symmetric about the vertical axis, the flow inside the alcove is asymmetric in the flow direction, and does not follow the interface near the ceiling. 

\begin{figure}
\centering
\includegraphics[width=\textwidth]{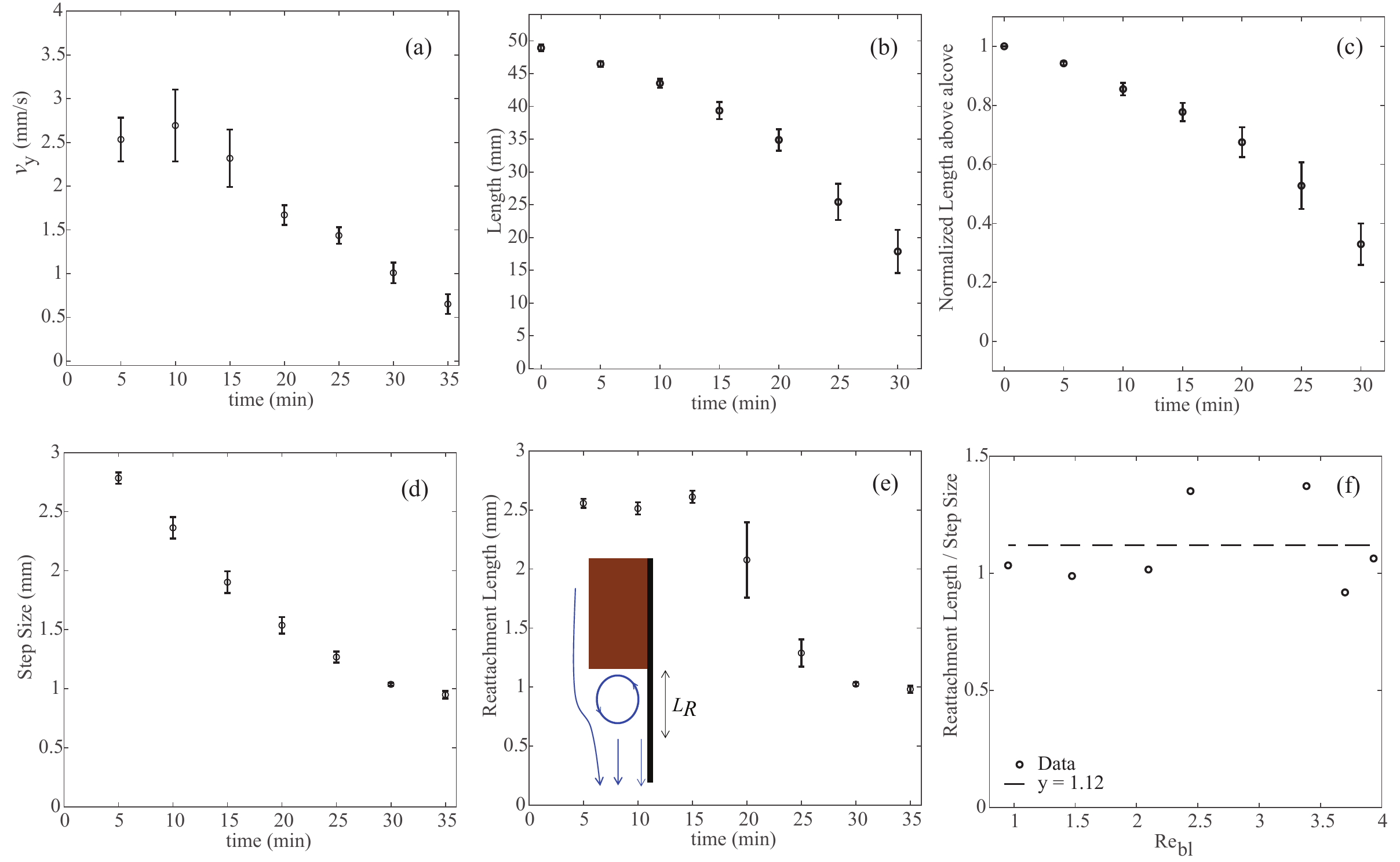}
\caption{(a) Average downward velocity of the gravity current along the dissolving vertical surface just above the alcove over time. The data is averaged over 5 examples and error bars corresponds to the root mean square deviations from the mean. (b) The total length of the interface decreases over time as the solid dissolves leading to the decrease in gravity current speed. (c) The fraction of the dissolving solid above the alcove also decreases proportionately. (d) The thickness of the dissolving solid above the alcove corresponds to the step size through which the gravity current passes as it enters the alcove region. (e) The reattachment length $L_R$ decreases as the step size decreases over time. Inset: Sketch of the flow and the corresponding $L_R$. (f) The ratio of the reattachment length to the step size remains approximately constant as a function of boundary layer Reynolds number calculated over the duration of the dissolution. Re$_{bl}$ remains moderate over the entire duration.}
\label{fig:flow-analy}
\end{figure}

To obtain a full picture of the flow, we visualize a central vertical thin ($\approx 100\,\mu$m) section through the alcove, by swapping the position of the camera and laser sheet to be from the side and front, respectively. A movie of the flow can be found in the supplementary documentations~\cite{sup-doc}. Figure~\ref{fig:flow}(c) then shows the flow visualized using the tracers over a 2\,second time interval in the orthogonal plane. A further zoomed in side view of the region near the ceiling is shown in Fig.~\ref{fig:flow}(d), where the tracers tracked over time are also shown. It is evident that there is clear flow separation as the gravity driven flow detaches from the surface at the edge where the vertical surface meets the ceiling of the alcove before reattaching to the back wall some distance down (see Fig.~\ref{fig:flow}(e)). 

From these complementary views, we conclude that the flow has a rich three dimensional structure with a predominantly downward flow which detaches at the alcove ceiling and reattaches inside the back of the alcove before flowing down and out.  Further, two sets of circulating currents or vortices are present inside the alcove. One gives rise to a return flow upwards in front of the alcove face away from the interface itself. The another, which is located between the ceiling and where the flow detaches from the ceiling, gives rise to flow which is directed outward along the alcove ceiling. 

The velocity $v_y$ along the interface was obtained by measuring the average length of the 10 longest streaks by eye from movies recorded at 24 frames per second in one minute time intervals over the course of the entire interface dissolution, and further averaging over 3 trails. These speeds in the flowing layer were observed to be right adjacent to the dissolving interface and are no more than a fraction of millimeter in width.  The measured speeds are plotted in Fig.~\ref{fig:flow-analy}(a), and observed to be,  initially, approximately 2.5 mm/s and then decrease over time. We attribute this decrease in speed to a finite size effect. As the interface dissolves, its total length inside the mold decreases after a few minutes because of relatively faster dissolution of the solid at the top and bottom edge of the mold interface (see Fig.~\ref{fig:flow}(a)). We quantify this by plotting the total length of the dissolving solid interface through the center in Fig.~\ref{fig:flow-analy}(b), and the fractional length dissolving solid interface above the alcove in Fig.~\ref{fig:flow-analy}(c). Both can be observed to decreases similarly, and one-to-one correspondence can be noted with the decreases in the flow speed.  

To understand this, we can note that the speed of the currents are determined by gravity acting on the solute concentrated solution, and the viscous drag of the flow across the interface and through the bath. Because the dissolving part of the interface decreases while the effective length of the recirculating flows through the bath remains unchanged, this can give rise to an overall slowing down of the flow as observed here.  The slowing down of the flow due to finite size effects leads to a decrease in alcove ceiling recession as can be noted in the overlap of the contour lines at later times in Fig.~\ref{fig:vertical}(a). Thus, the overall flow across the entire interface and the dissolution at the ceiling due to density-inversion, are both important to determining the overall evolution of the dissolving interface. 

The Reynolds number of the flow is given by ${\rm Re} = {\rho \, U l}/{\mu}$, where $\rho$ is the density of the solution, $\mu$ the viscosity, $U$ is the velocity scale, and $l$ is a length scale. Then, assuming $\mu > 8.90 \times 10^{-4}$\,Pa\,s, the viscosity of water at $25^\circ$C, and {\color{black} $l \sim H = 0.4$\,cm}, we have Re $< 2.4$. Thus, the boundary layer flows in our experiments are in the low-Re laminar flow regime, as also confirmed by tracer motion near the dissolving surface in the movies. At similar low Re numbers, vortices have been reported in the case of flow past a back step~\cite{Matsui75}, and is a result of the momentum of the fluid as it moves past the sharp discontinuity. To quantify the detachment of the flow, we measure the reattachment length $L_r$ from where the flow moves downward next to the wall to the ceiling as shown in inset to Fig.~\ref{fig:flow-analy}(e). Figure~\ref{fig:flow-analy}(d) and (e) show that the step length and the reattachment length of the flow is observed to decrease over time as the solid dissolves. And, the ratio of the reattachment length and the step size is observed to be roughly similar and constant over the duration of the dissolution as the Reynolds number of the boundary layer flow ${\rm Re}_{\rm bl}$ decreases slowly (see Fig.~\ref{fig:flow-analy}(f)). {\color{black} In calculating ${\rm Re}_{\rm bl}$, we have assumed that the relevant density and viscosity of the fluid is that of water, the length scale corresponding to the boundary layer is given by the fast moving layer in Fig.~\ref{fig:flow}(e), which is of order 100\,microns, and the flow speed $U$ is given by $v_y$ plotted in Fig.~\ref{fig:flow-analy}(a).}     

Combining our observations on the evolution of the alcove shape and gravity currents, one can surmise that the  alcove shapes evolve because of the two different flows at the ceiling and on the other sides of the alcove.  The flow separation and the counter-rotating vortex lead to a boundary layer which preserves the sharp discontinuity as the ceiling dissolves at a rate set largely by the the solutal Rayleigh-B\'enard instability that causes the ceiling to grow laterally.  Whereas, the flow does not separate from the boundary near the sides and the floor giving rise to dissolution rates set there by the flow tangential to the surface at those locations. This gives rise to the evolution of the surface which round out over time giving rise to a gradual change in slopes. {\color{black} Because the dissolution increases with tangential flow, the shapes can be expected to elongate along the flow direction. Hence this flow can give rise to the smooth conical surfaces below the ceiling which elongate over time. }

\section{Dissolution of Smooth and Rough Interfaces} 
\label{app:smooth}
\begin{figure}
\centering
\includegraphics[width=0.45\textwidth]{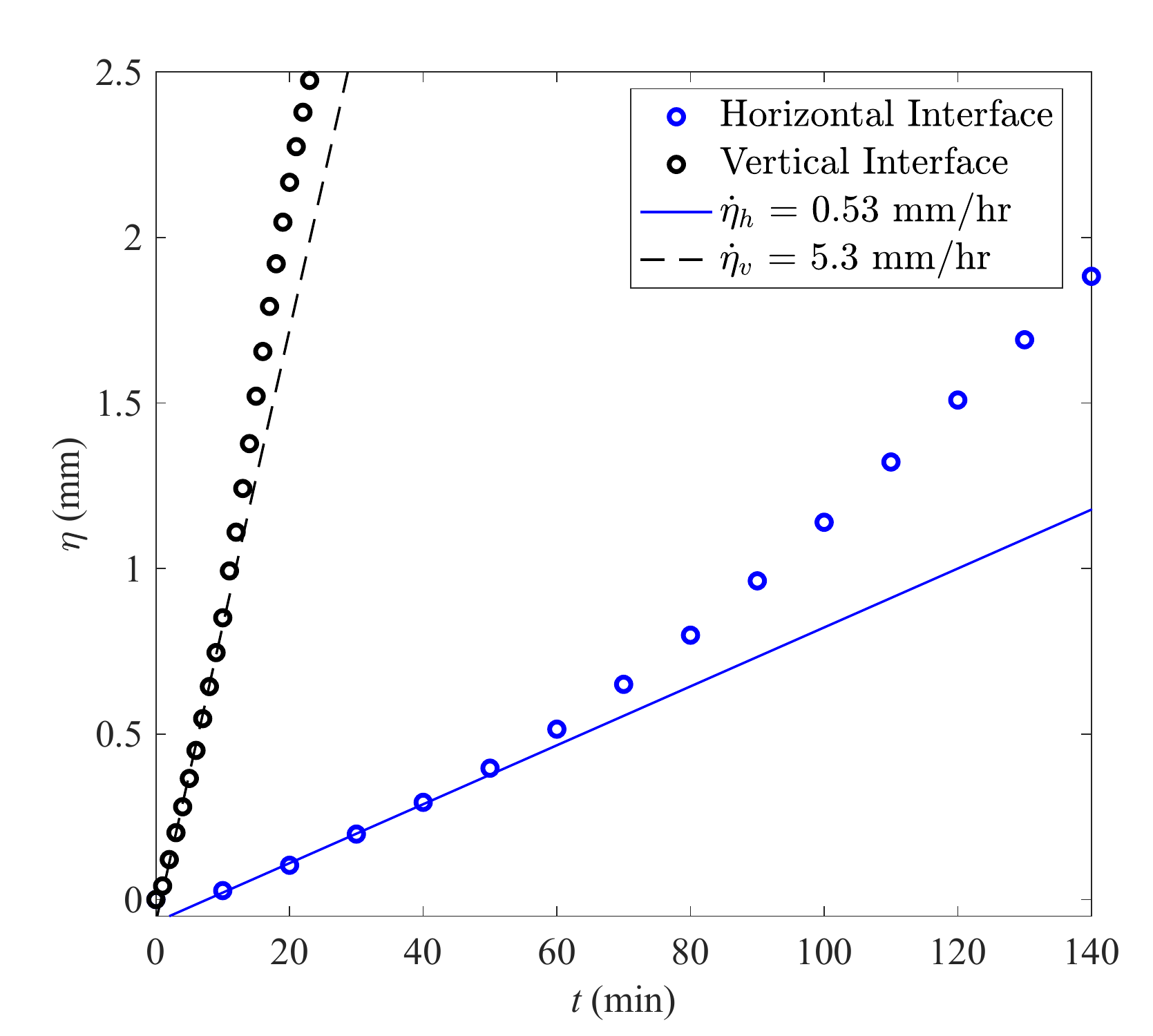}
\caption{The average thickness of the solid $\eta$ measured over time $t$ for the two orientations ($Q_{imp} = 0$\,mL/hr). The recession rates $\dot{\eta}$ are approximately 0.53\,mm/hr and 5.3\,mm/hr, respectively.}  
\label{fig:Fig3}
\end{figure}

In complementary experiments we examined the dissolution of solids with sufficiently smooth surfaces to see if the dissolving interface is indeed stable under otherwise similar preparation and environmental conditions. Solid blocks were prepared to be free of large bubble defects, and studied while the dissolving interface is oriented parallel and perpendicular with respect to gravity. No alcoves were observed to grow and the surfaces remained smooth and planar as they dissolve. 

Fig.~\ref{fig:Fig3}(a) shows a plot of the measured dissolved thickness layer $\eta(t)$ corresponding to the two orientations. The recession rate $\dot{\eta}$ is obtained using the slope of the dashed line fitted to the initial increase of $\eta(t)$ and is approximately 5.3\,mm/hr and 0.53\,mm/hr in the vertical and horizontal case, respectively. Thus, the dissolution is clearly greater in the case of the vertical orientation 
because convective currents which take away dense solution in the boundary layer are greater compared to the horizontal case. While dissolution driven current clearly occurs when the interface is vertical, driving the rapid descent of the denser solution, a slower circulation current is set up as well in the horizontal case. This occurs because of the finite size of the dissolving surface compared to the bath. Thus, as the sucrose diffuses up, the denser fluid slowly moves to the sides of the bath container as the interface is of finite size relative to the bath container. Thus, even in the horizontal interface, dissolution is not diffusion limited but assisted by convection, albeit much less so compared with a vertical interface. 

We then tested the fact that {\it indentations} of a critical size are needed to observe the growth of alcoves, and not just heterogeneity, by adding glass beads with a mean diameter of 0.2\,mm to the solid. This was accomplished by preparing the {\color{black} hot sucrose liquid} with the same recipe as noted earlier, and by adding beads corresponding to a volume fraction $V_f = 0.16$ as it sets. We find that although the glass beads protrude and peel off as the solid dissolves while the interface is oriented vertically, no alcoves form except where trapped air bubbles are present which were of diameter $d_b \gtrsim 0.63$\,mm. Alcoves were then observed to develop with inverted triangular shapes and to form at rates  similar to those observed without the beads added~\cite{sup-doc}.  It can be noted that this minimal defect size for alcove development is close to the critical size $\delta_c \approx 0.40$\,mm needed to trigger the solutal Rayleigh-Bénard instability according to Eq.~\ref{Eq1}.  Thus, indentations caused by trapped bubbles are apparently sufficiently great that the solutal Rayleigh-B\'enard instability can develop. The roughness and protrusions due the additional granular phase in these experiments appear to be unimportant to the development of the overall alcove shapes.  This data shows the robustness of the alcove feature even in the presence of additional non-dissolving solids as is often the case in natural sediments.

\section{Conclusions}

We have demonstrated that alcoves can develop on vertical solid-fluid interfaces as the solid phase dissolves when indentations of a critical size are present on the surface.  Unlike previous demonstrations of scallop-like indentations which cover the undersides of dissolving surfaces, these features are isolated, and small perturbations and protrusions are polished away as the interface dissolves. The alcoves are shown to develop starting at inverted surfaces (or ceilings) of sufficient size  as they dissolve leading to a boundary layer with a higher density compared to the solution below. From the observation of plumes descending from the ceiling, their spacing, and the observed recession rate of the ceiling, we deduce that the instability occurs when the boundary layer exceeds a length scale set by the critical Rayleigh number  corresponding to mixed boundary conditions. Thus, we identify the mechanism which leads to the formation of alcoves as being driven by the solutal Rayleigh-B\'enard instability as a result of the balance between the diffusion of the solute and the buildup of boundary layer density due to dissolution. 

By visualizing the fluid phase near the dissolving interface, we demonstrate that a gravity current develops as the density of solution increases with solute concentration along the entire dissolving interface and not just at the ceiling. As the fluid sinks, fresh fluid from the bath replaces it, and thus a rapid gravity current is setup moving downwards along the interface. We show that the boundary layer flow is in the low-Reynolds number laminar regime, and a boundary flow separation occurs at the edge where the vertical interface meets the ceiling of the indentations which enforces the discontinuous change in slope at the leading edge of the indentation.  The resulting dissolution as a result of the solutal Rayleigh-B\'enard instability, and the direction of the vortex which develops past the ceiling inside the indentation, appear to work in concert to maintain the sharp edge and widening of the ceiling as it continues to retreat upwards along the vertical interface. The continuing fluid flow down the side and floor of the indentation remain attached to the surface and serve to smooth sharp edges downstream leading to a gradual change in slope. The combination of density inversion instability at the ceiling and the shaping of the dissolving surfaces by the boundary layer flow result in the triangular shaped alcove with a wide ceiling and a sloping back and floor. The evolution of these shapes is found to be robust even when additional heterogeneity in the form of non-dissolving glass beads are present in the solid phase.  

Although the flow geometry is quite different than in confined triangular cavities that occur {\it inside} salt deposits studied by Gechter, et al.~\cite{Gechter2008}, an analogy can be observed. The appearance of dissolution at the ceiling due to the solutal Rayleigh-B\'enard instability leads to an efficient transfer of dissolved solute to regions where flow is rapid in turn giving rise to the triangular shapes. The flows in and out the alcove are however different than those in the formation of conical enclosed cavities in salt deposits. The rapid gravity currents over the entire open interface further shape the dissolving boundary leading to a recession of the alcove floor with a slope downwards polished by the attached boundary layer flow consistent with recent studies of dissolving pinnacles~\cite{Wykes2018,Pegler2021}.  Further modeling is needed to draw quantitative comparisons between features observed in these various interfaces shaped by dissolution and flow. 

Finally, we note that our two-phase physical model system is highly simplified compared with alcoves observed in nature. Natural cliffs have far more complex rock chemistry and heterogeneity, and can experience a wide range of physical weathering from temperature variation to rain and wind besides chemical and biological weathering. Dissolution by surface runoffs fed by rainfall  may play a role in the initiation of surface patterns~\cite{Guerin2020}. Nonetheless, recent experiments with dissolving solids in an aqueous bath demonstrate that convective dissolution reproduce the shape of the pinnacles created by the rain in Karst regions~\cite{Huang2020}. {\color{black} In these studies, as in ours, dissolving sugar results in considerable variation in viscosity and diffusivity, which is not the case in typical rock dissolution.} The universality of the overall alcove shapes that evolve, largely independent of initial conditions in our system point to the underlying robustness of such features in dissolving solids shaped by gravity driven boundary layer flows. Exploring these connections more deeply remains an avenue for further research.

\appendix

\section{Imaging}
\label{sec:img}
\begin{figure}
\centering
\includegraphics[width=0.95\textwidth]{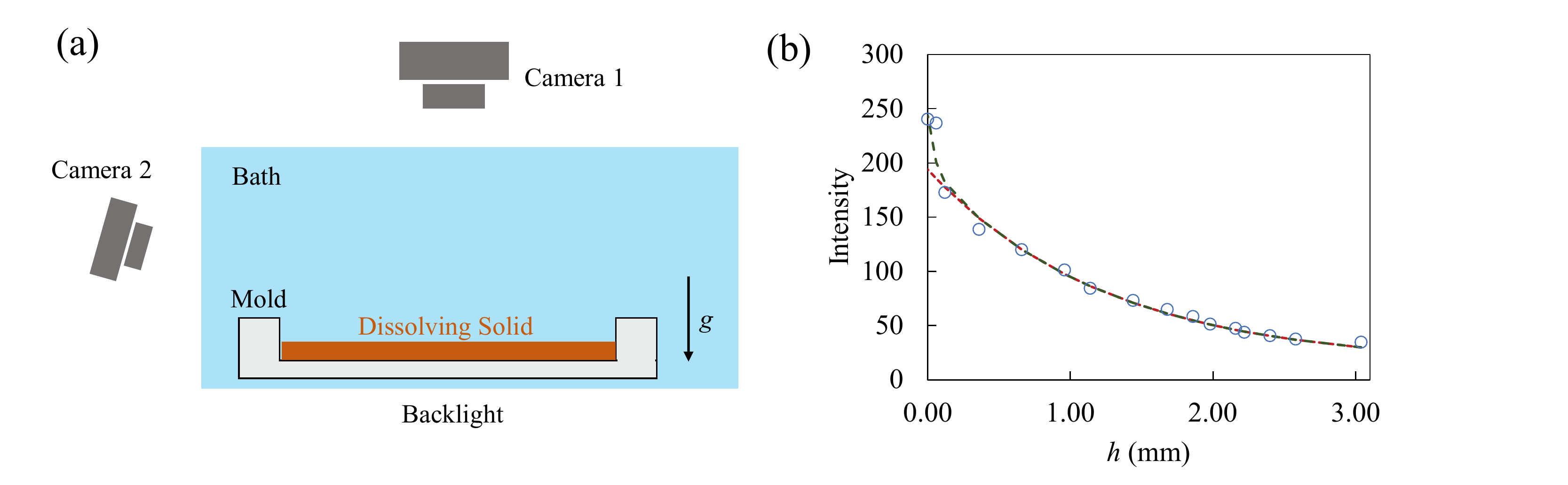}
\caption{(a) Schematic of the system used for calibrating light intensity to solid thickness $h$. (b) Light intensity measured as a function of solid thickness $h$. An exponential fit is found to describe the data for $h > 0.1$\,mm.}
\label{fig:calib}
\end{figure}

To map the observed intensity to the height of the dissolving solid, we prepared a system with two cameras configured as shown in Fig.~\ref{fig:calib}(a) to simultaneously acquire images. The solid in mold is placed horizontally in a bath of distilled water. As the solid dissolves the concentrated mixture settles on top of the undissolved solid. This mixture is displaced using a Pasteur pipette filled with distilled water to keep the bath water near the interface from changing color due to the high concentration of the dissolved solute. Images of the cleaned surface are then acquired every 5 minutes. Camera 1 placed directly above the dissolving solid captures the intensity corresponding to the solid sugar. 
Camera 2 placed at a grazing angle captures the wall of the mold as it becomes visible. Since we know the dimensions of the mold when filled with the dissolving solid and without any dissolving solid present, we use this configuration to determine the actual height of the mold wall becoming visible. Consequently, we determine the height of the dissolving solid. Since these are imaged simultaneously, we can compare the height of the solid at a given time with the average intensity observed at that same time. This data is plotted in Fig.~\ref{fig:calib}(b), and is described by the function 
$I(h) = 14 + 180 \exp(-0.8\,h)$, where the thickness of the block is greater than 0.1\,mm, {\color{black} consistent with Beer-Lambert law with a uniform light attenuation through the solid. }

\section{Horizontal Interface}
\label{sec:hori}
\begin{figure}
\centering
\includegraphics[width=0.85\textwidth]{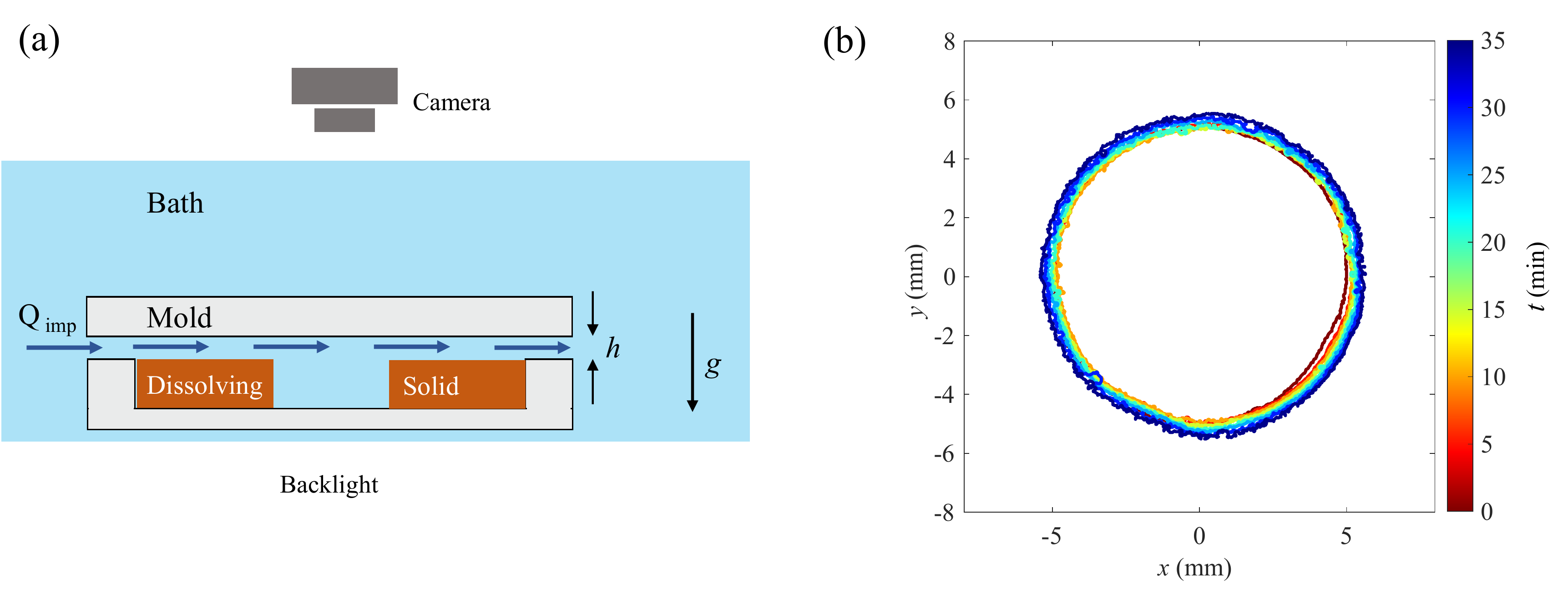}
\caption{(a) Schematic of the experimental system used to measure the dissolution of the interface when placed horizontally in the bath. A syringe pump is used to impose a flow across the interface with a prescribed rate. (b) Contour plot of the cavity development as a function of time corresponding to $Q_{imp} = 100$\,mL/h. The cavity remains circular.   
}
\label{fig:horcomp}
\end{figure}

To understand the conditions under which these features grow, we also examine the dissolution dynamics with a similarly prepared interface, while placed horizontally as shown schematically in Fig.~\ref{fig:horcomp}(a). A flow of $Q = 100$\,ml/h, corresponding to an average speed of 0.7 mm/s initially, similar in order of magnitude as those which arise near a vertical interface, 
is imposed across the horizontal interface to create a flow across the interface and flush out the dissolved sugar. The dissolution occurs relatively faster, in 2\,hours (135 minutes), compared to when no flow is imposed. Nonetheless the cavity evolves uniformly. As can be seen from the contour line plots in Fig.~\ref{fig:horcomp}(b), and the movie in the supplementary documentation~\cite{sup-doc} corresponding to the entire 2-hour duration, the interface dissolves more or less uniformly, and the cavity does not grow significantly in diameter even as the entire solid dissolves. Thus, it is not just the presence of surface perturbation which is important to the formation of the triangular shaped alcoves, but also the orientation of the interface.

\begin{acknowledgments}   
This research is supported by the U.S. Dept. of Energy, Office of Science, Basic Energy Sciences Grant DE-SC0010274. 
\end{acknowledgments}  

\bibliographystyle{unsrt}

\begin{thebibliography}{10}

\bibitem{Fredd1998}
Christopher~N. Fredd and H.~Scott Fogler.
\newblock The kinetics of calcite dissolution in acetic acid solutions.
\newblock {\em Chemical Engineering Science}, 53(22):3863--3874, 1998.

\bibitem{Ortoleva1994}
Peter~J Ortoleva.
\newblock {\em Geochemical self-organization}.
\newblock New York : Oxford University Press ; Oxford : Clarendon Press, 1994.

\bibitem{Malin2000}
Michael~C. Malin and Kenneth~S. Edgett.
\newblock Evidence for recent groundwater seepage and surface runoff on mars.
\newblock {\em Science}, 288(5475):2330--2335, 2000.

\bibitem{Meakin2010}
Paul Meakin and Bjørn Jamtveit.
\newblock Geological pattern formation by growth and dissolution in aqueous
  systems.
\newblock {\em Proc. R. Soc. A}, 466:659–694, 2010.

\bibitem{Kudrolli2016}
Arshad Kudrolli and Xavier Clotet.
\newblock Evolution of porosity and channelization of an erosive medium driven
  by fluid flow.
\newblock {\em Phys. Rev. Lett.}, 117:028001, 2016.

\bibitem{Jerolmack2019}
D.J. Jerolmack and K.E. Daniels.
\newblock {Viewing Earth’s surface as a soft-matter landscape}.
\newblock {\em Nature Reviews Physics}, 1:716--730, 2019.

\bibitem{Thomas1968}
D.~G. Thomas and R.~A. Armistead.
\newblock Concentration-gradient-driven convection: Experiments.
\newblock {\em Science}, 160(3831):995, 1968.

\bibitem{Sullivan96}
Timothy~S. Sullivan, Yuanming Liu, and Robert~E. Ecke.
\newblock Turbulent solutal convection and surface patterning in solid
  dissolution.
\newblock {\em Phys. Rev. E}, 54:486--495, 1996.

\bibitem{Gechter2008}
Daniel Gechter, Peter Huggenberger, Philippe Ackerer, and H.~Niklaus Waber.
\newblock Genesis and shape of natural solution cavities within salt deposits.
\newblock {\em Water Resources Research}, 44(11), 2008.

\bibitem{Huang2015}
Jinzi~Mac Huang, M.~Nicholas~J. Moore, and Leif Ristroph.
\newblock {Shape dynamics and scaling laws for a body dissolving in a fluid
  flow}.
\newblock {\em J. Fluid Mech.}, 765:R3--1--9, 2015.

\bibitem{Nakouzi2015}
Elias Nakouzi, Raymond~E. Goldstein, and Oliver Steinbock.
\newblock {Do Dissolving Objects Converge to a Universal Shape?}
\newblock {\em Langumir}, 31:4145--4150, 2015.

\bibitem{Cohen2016}
Caroline Cohen, Michael Berhanu, Julien Derr, and Sylvain Courrech~du Pont.
\newblock Erosion patterns on dissolving and melting bodies (2015 gallery of
  fluid motion).
\newblock {\em Phys. Rev. Fluids}, 1:050508, 2016.

\bibitem{Huang2020}
Jinzi~Mac Huang, Joshua Tong, Michael Shelley, and Leif Ristroph.
\newblock {Ultra-sharp pinnacles sculpted by natural convective dissolution}.
\newblock {\em PNAS}, 117:23339--23344, 2020.

\bibitem{Wykes2018}
Megan S.~Davies Wykes, Jinzi~Mac Huang, George~A. Hajjar, and Leif Ristroph.
\newblock {Self-sculpting of a dissolvable body due to gravitational
  convection}.
\newblock {\em Physical Review Fluids}, 3:043801--1--18, 2018.

\bibitem{Pegler2020}
Samuel~S. Pegler and Megan S.~Davies Wykes.
\newblock {Shaping of melting and dissolving solids under natural convection}.
\newblock {\em J. Fluid Mech.}, 900:A35, 2020.

\bibitem{Cohen2020}
Caroline Cohen, Michael Berhanu, Julien Derr, and Sylvain Courrech~du Pont.
\newblock Buoyancy-driven dissolution of inclined blocks: Erosion rate and
  pattern formation.
\newblock {\em Phys. Rev. Fluids}, 5:053802, 2020.

\bibitem{bushuk_holland_stanton_stern_gray_2019}
Mitchell Bushuk, David~M. Holland, Timothy~P. Stanton, Alon Stern, and Callum
  Gray.
\newblock Ice scallops: a laboratory investigation of the ice–water
  interface.
\newblock {\em Journal of Fluid Mechanics}, 873:942–976, 2019.

\bibitem{sup-doc}
See Supplemental Material at [URL will be inserted by publisher] for movies of
  the dissolution dynamics.

\bibitem{Garner1961}
F.~H. Garner and J.~M. Hoffman.
\newblock Mass transfer from single solid spheres by free convection.
\newblock {\em A.I.Ch.E. Journal}, 7(1):148, 1961.

\bibitem{Pegler2021}
Samuel~S. Pegler and Megan S.~Davies Wykes.
\newblock {The convective Stefan problem: shaping under natural convection}.
\newblock {\em J. Fluid Mech.}, 915:A86, 2021.

\bibitem{Chandrasekhar}
Subrahmanyan Chandrasekhar.
\newblock {\em Hydrodynamic and Hydromagnetic Stability}.
\newblock Clarendon Press, Oxford, 1961.

\bibitem{Philippi2019}
Julien Philippi, Michael Berhanu, Julien Derr, and Sylvain Courrech~du Pont.
\newblock Solutal convection induced by dissolution.
\newblock {\em Phys. Rev. Fluids}, 4:103801, 2019.

\bibitem{Matsui75}
T.~Matsui, M.~Hiramatsu, and M.~Hanaki.
\newblock Separation of low reynolds number flows around a corner.
\newblock {\em Symposia on Turbulence in Liquids}, 30, Sept. 1975.

\bibitem{Guerin2020}
Adrien Guerin, Julien Derr, Sylvain~Courrech du~Pont, and Michael Berhanu.
\newblock {Streamwise Dissolution Patterns Created by a Flowing Water Film}.
\newblock {\em Phys. Rev. Lett.}, 125:194502--1--6, 2020.

\end{thebibliography}

\end{document}